\documentclass[aps,prd,reprint,superscriptaddress,nofootinbib]{revtex4-1}
\usepackage{blindtext}
\usepackage{hyperref}
\usepackage{tikz}
\usepackage{amsmath,amssymb}
\usepackage{mathtools}
\usepackage{hyperref}
\usepackage{undertilde}
\usepackage{graphicx}
\usepackage{hyperref}

\begin{document}
\title{Transport coefficients from in medium quarkonium dynamics}

\author{Nora Brambilla}
\email{nora.brambilla@tum.de}
\affiliation{Physik Department, Technische Universit{\"a}t M{\"u}nchen, 85748 Garching, Germany}
\affiliation{Institute for Advanced Study, Technische Universit{\"a}t M{\"u}nchen, Lichtenbergstrasse 2 a, 85748 Garching, Germany}

\author{Miguel A. Escobedo}
\email{miguelangel.escobedo@usc.es}
\affiliation{Instituto  Galego  de  F\'{i}sica  de  Altas  Enerx\'{i}as (IGFAE), Universidade  de  Santiago  de  Compostela, Galicia-Spain}
\affiliation{Department of Physics, P.O. Box 35, FI-40014 University of Jyv\"askyl\"a, Finland}

\author{Antonio Vairo}
\email{antonio.vairo@tum.de}
\affiliation{Physik Department, Technische Universit{\"a}t M{\"u}nchen, 85748 Garching, Germany}

\author{Peter Vander Griend}
\email{vandergriend@tum.de}
\affiliation{Physik Department, Technische Universit{\"a}t M{\"u}nchen, 85748 Garching, Germany}

\begin{abstract}
	The in medium dynamics of heavy particles are governed by transport coefficients.
	The heavy quark momentum diffusion coefficient, $\kappa$, is an object of special interest in the literature,
	but one which has proven notoriously difficult to estimate,
	despite the fact that it has been computed by weak-coupling methods at next-to-leading order accuracy, 
	and by lattice simulations of the pure SU(3) gauge theory.
	Another coefficient, $\gamma$, has been recently identified. It can be understood as the dispersive counterpart of $\kappa$.
	Little is known about $\gamma$.
	Both $\kappa$ and $\gamma$ are, however, of foremost importance in heavy quarkonium physics
	as they entirely determine the in and out of equilibrium dynamics of quarkonium in a medium, 
	if the evolution of the density matrix is Markovian, and the motion, quantum Brownian; 
	the medium could be a strongly or weakly coupled plasma.
	In this paper, using the relation between $\kappa$, $\gamma$ and the quarkonium in medium width and mass shift respectively,
	we evaluate the two coefficients from existing 2+1 flavor lattice QCD data.
	The resulting range for $\kappa$ is consistent with earlier determinations,
	the one for $\gamma$ is the first non-perturbative determination of this quantity.
\end{abstract}

\maketitle

\section{Introduction}
\label{sec:intro}
Heavy quarkonium has long been theorized to serve as a probe of the medium formed in heavy ion collisions 
with the purpose to detect a new state of matter, the quark gluon plasma (QGP)~\cite{Matsui:1986dk}.
In turn, the study of the QGP offers a unique window on the universe at an early time.

In this paper, we focus on the out of equilibrium dynamics of heavy quarkonium in the medium. Our aims are twofold.
Making use of recent results~\cite{Brambilla:2016wgg,Brambilla:2017zei}, we further elaborate on the out of equilibrium dynamics
under the assumptions that the evolution of the quarkonium density is Markovian, and the motion, quantum Brownian.
We emphasize the relation that exists, under these conditions, between the quarkonium dynamics and the transport coefficient~$\kappa$, 
describing the momentum diffusion of a heavy quark in a medium, and $\gamma$, the dispersive counterpart of~$\kappa$.
Finally, exploiting this relation, and under the conditions of its validity, 
we provide a method for extracting the coefficients $\kappa$ and $\gamma$ from existing 2+1 flavor lattice QCD data.
A recent work on the extraction of heavy quark transport coefficients is Ref.~\cite{Cao:2018ews}.

The remainder of the paper is structured as follows.
In Section~\ref{sec:background}, we summarize some recent progress in the study of the out of equilibrium dynamics of heavy particles
and, in particular, heavy quarkonium in a medium, supplying the relevant background for the results to follow. 
Section~\ref{sec:theorycalc} contains the evolution equations for the heavy quarkonium density matrix in a medium 
under the condition of Markovianity and quantum Brownian motion.
In this section, we also relate the transport coefficients $\kappa$ and $\gamma$ with the quarkonium in medium width and mass shift respectively.
In Section~\ref{sec:numerical_calculations}, we use existing lattice data to assign numerical values to both coefficients. We conclude in Section~\ref{sec:conclusion}.
Technical details can be found in Appendix~\ref{appendix}.

\section{Theoretical background}\label{sec:background}

\subsection{Heavy quarkonia via Lindblad equation: $\kappa$ and $\gamma$}
\label{subsec:lindblad}
A recent body of work has sought to model heavy quarkonium evolution in the medium formed in heavy ion collisions using the formalism of open quantum
systems~\cite{Borghini:2011ms,Akamatsu:2014qsa,Brambilla:2016wgg,Kajimoto:2017rel,Brambilla:2017zei,Blaizot:2017ypk,Akamatsu:2018xim,Yao:2018nmy,Yao:2018sgn}.
For a general review of open quantum systems, we direct the reader to~\cite{Breuer:2002pc}. 
Quarkonium serves as the {\em system}, and the medium, which can be a QGP, as the {\em environment}.  

The system, quarkonium, is characterized by at least three energy scales:
the mass $M$ of the heavy quark, the inverse of the Bohr radius, $a_0$, and the binding energy $E$.
These energy scales, quarkonium being a non-relativistic bound state, are hierarchically ordered: $M \gg 1/a_0 \gg E$.
We identify the inverse of $E$ with the {\em intrinsic time scale of the system}: $\tau_{S}\sim 1/E$.

The environment, the medium, may be characterized by several energy scales.
We will assume just one single energy scale, $\pi T$. 
We identify the inverse of $\pi T$ with the {\em correlation time of the environment}: $\tau_{E}\sim 1/(\pi T)$.
If the medium is in thermal equilibrium, or locally in thermal equilibrium, we may understand $T$ as the temperature.
Since we do not exploit any further, possible, hierarchy among the energy scales of the medium, 
we are indeed considering that the medium may be strongly coupled. 
For instance, if the medium is a strongly coupled QGP in local thermal equilibrium, 
then $\pi T$ is of the same order as the Debye mass, $m_D \sim gT$, and of the same order as the magnetic screening mass, $m_M \sim g^2T$.

The evolution of the system in the environment is characterized by a {\em relaxation time} $\tau_R$.
We assume that the quarkonium is {\em Coulombic}, which applies to the charmonium and bottomonium ground states.
This requires that 
\begin{equation}
\frac{1}{a_0} \gg \pi T, \Lambda_{\text{QCD}}\,,
\label{ass-coul}
\end{equation}
and that in medium and non-perturbative corrections to the Coulomb potential are subleading.\footnote{The temperature of the medium formed in a heavy ion collision at the LHC ranges from approximately 475 MeV down to the freeze-out temperature giving a maximum of $\pi T$ of approximately 1.5 GeV in the initial stages of the collision.  In Sec.~\ref{sec:numerical_calculations}, for the $\Upsilon(1S)$ state, we calculate $1/a_{0}\approx 1.5$ GeV. As the medium expands rapidly in the initial stages of the thermal evolution, it quickly cools to lower temperatures, and for the lowest lying bottomonium states,we expect Eq.~(\ref{ass-coul}) to hold for all but the earliest times.  As the $J/\psi$ has a significantly larger radius than the $\Upsilon(1S)$, this relation will hold at lower temperatures.  In Sec.~\ref{sec:numerical_calculations} for the $J/\psi$, we calculate $1/a_{0}\approx 0.84$ GeV giving a range of validity up to temperatures of approximately 250 MeV.}  Furthermore, we also assume that 
\begin{equation}
\pi T \gg E\,.
\label{ass-Brown}
\end{equation}
Under these assumptions the relaxation time is given by the inverse of the self-energy diagram shown in Fig.~\ref{fig:sigmas},
\begin{equation}
\tau_R \sim \frac{1}{\Sigma_s} \sim \frac{1}{a_{0}^{2}(\pi T)^{3}}.
\end{equation}
In the case of a weakly coupled medium, the relaxation time may be enhanced by a factor $1/g^2(T)$, $g$ being the QCD gauge coupling. 

Under the Coulombic assumption \eqref{ass-coul}, it follows that 
\begin{equation}
\tau_{R}\gg \tau_{E},
\label{RggE}
\end{equation}
which is a necessary condition for the system to be insensitive to the initial condition of the environment and, therefore, to show a {\em Markovian evolution}.
Moreover, from \eqref{ass-Brown} it follows that 
\begin{equation}
\tau_{S}\gg \tau_{E}.
\label{RggS}
\end{equation}
This qualifies the regime of the quarkonium in the medium as {\em quantum Brownian motion}~\cite{Akamatsu:2014qsa}.   

In~\cite{Brambilla:2016wgg,Brambilla:2017zei} it has been shown that under the Coulombic, Eq.~\eqref{ass-coul} or Eq.~\eqref{RggE}, and 
the Brownian motion assumption, Eq.~\eqref{ass-Brown} or Eq.~\eqref{RggS}, the evolution equation for the density matrix, $\rho$, of the 
heavy quark-antiquark system can be written in the {\em Lindblad form}~\cite{Lindblad:1975ef,Gorini:1975nb}:
\begin{equation}
\frac{\mathrm{d}\rho}{\mathrm{d}t}=-i[H,\rho]+\sum_{n}\left( C_{n}\rho C_{n}^{\dagger}-\frac{1}{2}\left\{ C_{n}^{\dagger}C_{n},\rho \right\} \right),
\label{eq:lindblad}
\end{equation}
where $H$ is a Hermitian operator, and $C_{n}$ are known as {\em collapse operators}. 
These operators were computed in~\cite{Brambilla:2016wgg,Brambilla:2017zei}; 
we give the explicit expressions of $\rho$, $H$, and the $C_{n}$ in the following Eqs.~\eqref{densitymatrix} and~\eqref{eq:hamiltonian} to~\eqref{eq:collapse1} in Section~\ref{sec:theorycalc}.  
The operators $H$ and $C_{n}$ turn out to depend on only two transport coefficients, $\kappa$ and $\gamma$, that encode the entire in medium dynamics.
They are related to the real and imaginary parts of the heavy quarkonium self energy, $\Sigma_s$, with a rigorous exposition and derivation of $\kappa$ and $\gamma$ in the context of heavy quarkonium presented in Sec.~\ref{sec:theorycalc}.
In~\cite{Brambilla:2016wgg,Brambilla:2017zei} it was further recognized that $\kappa$ is in fact the {\em heavy quark momentum diffusion coefficient}, 
while $\gamma$ could be understood as its dispersive counterpart.
We elaborate more on $\kappa$ and its role in the in medium dynamics of heavy quarks in the following section.

\subsection{Heavy quarks via Langevin equation: $\kappa$}
\label{subsec:langevin}
The heavy quark momentum diffusion coefficient $\kappa$ is an object
of great interest in the literature~\cite{Moore:2004tg,CasalderreySolana:2006rq,CaronHuot:2008uh,CaronHuot:2009uh,Banerjee:2011ra,Francis:2015daa,Xu:2017obm} 
as it affects the momentum distribution of heavy flavor mesons measured in several experimental facilities~\cite{Adare:2010de,Akiba:2015jwa,Acharya:2017qps,Adamczyk:2017xur}.
It is a key component in understanding the heavy quark diffusion in a thermal medium in the framework of the {\em Langevin equations}.
Specifically, for a heavy quark of mass $M$ in a thermal medium at a temperature $T$, with $M\gg T$, 
the momentum of the heavy quark changes little over the characteristic time scale of the plasma 
due to random interactions with the medium constituents~\cite{Moore:2004tg}.  
This slow evolution due to uncorrelated interactions with the medium is described by the Langevin equations:
\begin{equation}\label{eq:langevin}
\frac{\mathrm{d}p_{i}}{\mathrm{d}t}=-\eta_{D}p_{i}+\xi_{i}(t), \qquad 
\langle \xi_{i}(t)\xi_{j}(t')\rangle=\kappa \delta_{ij}\delta(t-t'),
\end{equation}
where $p_{i}$ is the momentum of the heavy quark, $\eta_{D}$ is the {\em drag coefficient}, and $\xi_{i}$ encodes the random, 
uncorrelated interactions of the quark with the medium.  
Demanding that the system approaches thermal equilibrium entails an Einstein relation between the drag coefficient 
and the heavy quark momentum diffusion coefficient, i.e., $\eta_{D} = \kappa/(2MT)$. 
We thus see that the dynamics of the heavy quark in the thermal medium are governed by a single transport coefficient, 
namely the heavy quark momentum diffusion coefficient $\kappa$.

\subsection{Determining $\kappa$ and $\gamma$}
\label{subsec:relations}
In spite of the relevance of $\kappa$ in the theoretical description of heavy quark diffusion in a thermal medium, 
its calculation has proven arduous and a precise determination elusive; for recent reviews see~\cite{Xu:2018gux,Dong:2019byy,Dong:2019unq,Rapp:2018qla}.  
Calculations of $\kappa$ require a number of assumptions on the dynamics and initial conditions
of the medium along with its evolution and interaction with the heavy quark.  
Comparison with data would then allow for a discrimination among different assumptions and models.  
Large experimental uncertainties combined with subtle interactions among different assumptions have complicated the attempts to fix $\kappa$ reliably.  
The heavy quark momentum diffusion coefficient may be also determined by means of lattice simulations.
While the extraction from the spectral function of current-current correlators has turned out to be very difficult~\cite{Petreczky:2005nh},  
more recently, $\kappa$ has been related to the {\em spectral function} of the chromoelectric field correlator $\langle gE^{a,i}(t,\textbf{0})\,gE^{a,i}(0,\textbf{0}) \rangle$, 
$\rho_{\text{el}}$, in thermal QCD~\cite{CasalderreySolana:2006rq,CaronHuot:2009uh}. For definitions and details, see Appendix~\ref{appendix}.
The relation reads
\begin{equation}
\kappa =  \frac{T}{6N_c}\, \lim_{\omega\to 0} \frac{\rho_{\text{el}}(\omega)}{\omega},
\label{kappa-spectral}
\end{equation}
which constrains $\kappa$ to be positive. $N_c=3$ is the number of colors. 
Equation~\eqref{kappa-spectral} has allowed to determine $\kappa$, so far, on quenched lattices in thermal QCD  
for temperatures between $T_c$ and $2T_c$, $T_c$ being the crossover temperature to the QGP~\cite{Banerjee:2011ra,Francis:2015daa}.
Finally, an analytic, perturbative estimate of $\kappa$ up to next-to-leading order in the hard-thermal-loop effective theory 
appears to suffer from poor convergence~\cite{CaronHuot:2008uh}.  

In this paper, we determine $\kappa$ from the thermal decay width of a heavy quarkonium in a strongly coupled medium~\cite{Brambilla:2016wgg,Brambilla:2017zei}.
This determination uses a different observable, the quarkonium thermal width, a different set of assumptions, Eqs.~\eqref{ass-coul} (Coulombic bound state) and~\eqref{ass-Brown}
(quantum Brownian motion), and a different source of data, 2 flavor lattice QCD data from~\cite{Aarts:2011sm} and 2+1 flavor lattice QCD data from~\cite{Kim:2018yhk}.
Therefore, it is an independent determination with different systematic uncertainties, potentially competitive with other determinations.
We estimate our main sources of systematic uncertainties in this determination of $\kappa$ to be higher order corrections inherent in our effective field theory approach and the systematic uncertainties inherited from the specific lattice data used in our calculations. As discussed in Sec.~\ref{subsec:lindblad}, for bottomonium, we expect our hierarchy of scales in Eq.~(\ref{ass-coul}) to be fulfilled and these higher-order corrections to be small.

In contrast to the theoretical understanding of the role of $\kappa$ in the dynamics of heavy quarks in a thermal medium and the progress towards its calculation, comparatively little is known about $\gamma$. 
Since $\gamma$ may be understood as a correction to the heavy quark-antiquark potential, no similar object arises in the description of the in medium heavy quark dynamics. 
A proper definition relates $\gamma$ to the chromoelectric field correlator in such a way that it may be considered the dispersive counterpart of~$\kappa$~\cite{Brambilla:2016wgg,Brambilla:2017zei}. 
Like $\kappa$, $\gamma$ can be written in thermal QCD in terms of the chromoelectric spectral function $\rho_{\text{el}}$:
\begin{equation}
\gamma = - \frac{1}{3N_c} \int_0^\infty \frac{\mathrm{d}\omega}{2\pi} \, \frac{\rho_{\text{el}}(\omega)}{\omega}.
\label{gamma-spectral}
\end{equation}
A derivation of Eq.~\eqref{gamma-spectral} is in Appendix~\ref{appendix}.
Differently from $\kappa$, however, the coefficient $\gamma$ is a function of ${\rho_{\text{el}}(\omega)}/{\omega}$ over the whole spectrum of frequencies.
Since $\rho_{\text{el}}(\omega) \sim \omega^3$ for large frequencies, the above integral is ultraviolet divergent and needs to be regularized and renormalized.
The large frequencies behaviour of $\rho_{\text{el}}(\omega)$ is entirely given by the in vacuum ($T=0$) contributions.
These are known up to next-to-leading order~\cite{Burnier:2010rp}.
From Ref.~\cite{Burnier:2010rp} it also follows that the thermal part of $\gamma$ is finite.

Just as $\kappa$ is the parameter of central importance in the study of in medium heavy quarks, 
$\kappa$ and $\gamma$ appear to be the parameters of central importance in the study of the quantum Brownian motion of Coulombic quarkonia in a strongly coupled medium. 
In this paper, taking advantage of the relation between $\gamma$ and the quarkonium thermal mass shift in a strongly coupled medium~\cite{Brambilla:2016wgg,Brambilla:2017zei}, 
we determine the thermal part of $\gamma$ from the 2+1 flavor lattice QCD data of~\cite{Kim:2018yhk}.
The procedure will be similar to the one used to extract~$\kappa$, as well as the underlying assumptions~\eqref{ass-coul} and~\eqref{ass-Brown}.

\section{Quarkonium in the quantum Brownian regime}
\label{sec:theorycalc}
\begin{figure}[ht]
\includegraphics[width=\linewidth]{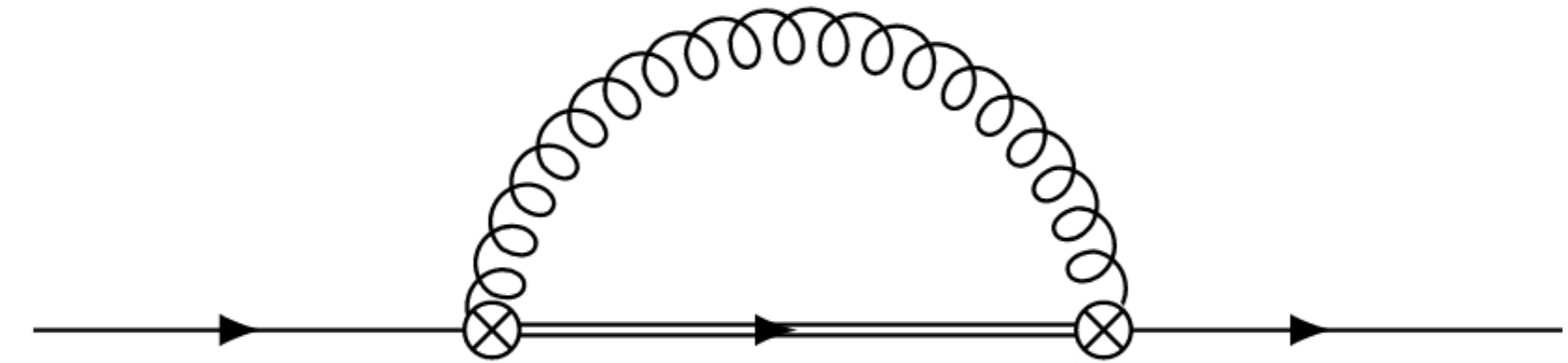}
\caption{A diagrammatic representation of the leading order color-singlet self-energy diagram, $\Sigma_{s}$, in pNRQCD. 
Single lines represent quark-antiquark color-singlet propagators, double lines quark-antiquark color-octet propagators, 
curly lines gluons, and crossed circles chromoelectric dipole vertices.}
\label{fig:sigmas}
\end{figure}

In~\cite{Brambilla:2017zei}, a set of master equations governing the time evolution of heavy quarkonium in a medium were derived. 
The equations follow from assuming the inverse Bohr radius of the quarkonium to be greater than the energy scale of the medium, Eq.~\eqref{ass-coul},
and model the quarkonium as evolving in the vacuum up to a time $t=t_{0}$, at which point interactions with the medium begin. 
The equations express the time evolution of the density matrices of the heavy quark-antiquark color singlet, $\rho_s$, and octet states, $\rho_o$,   
in terms of the color singlet and octet Hamiltonians, $h_s = {\bf p}^2/M - C_F\alpha_s/r + ...$ and 
$h_o = {\bf p}^2/M + \alpha_s/(2N_cr) + ...$,  and interaction terms with the medium, which, at order $r^2$ in the multipole expansion, 
are encoded in the self-energy diagram shown in Fig.~\ref{fig:sigmas}. These interactions account for the mass shift 
of the heavy quark-antiquark pair induced by the medium, its decay width induced by the medium, the generation of quark-antiquark color singlet states 
from quark-antiquark color octet states interacting with the medium and the generation of quark-antiquark color octet states
from quark-antiquark (color singlet or octet) states interacting with the medium. The color singlet and octet Hamiltonians, $h_s$ and $h_o$, 
describe particles of mass $M$ and momentum ${\bf p}$ interacting at a distance $r$ through a Coulomb potential; 
$C_F=(N_c^2-1)/(2N_c)=4/3$ is the Casimir of the fundamental representation, and $\alpha_{s}=g^2/(4\pi)$ is the strong coupling. 
The dots in our expressions of $h_s$ and $h_o$ stand for higher-order terms 
that are irrelevant for the present analysis. The effective field theory framework in which the non-relativistic heavy quark-antiquark dynamics 
can be systematically described in terms of quark-antiquark color singlet and octet fields, whose interactions with the medium are organized in powers 
of $1/M$ and $r$, is potential non-relativistic QCD (pNRQCD)~\cite{Pineda:1997bj,Brambilla:1999xf,Brambilla:2004jw}.
The leading order interaction between a heavy quark-antiquark field and the medium is encoded in pNRQCD in a chromoelectric dipole interaction,
which appears at order $r/M^0$ in the effective field theory Lagrangian.

Further assuming that any energy scale in the medium is larger than the heavy quark-antiquark binding energy,\footnote{
Note that, in pNRQCD, quark-antiquark pairs in a color octet configuration have energy ${\bf p}^2/M$, where ${\bf p}$ is a relative momentum of the	same order as $1/a_0$.  Hence the typical energy of the color octet pair is of order $E$, which is the binding energy of the color singlet pair.} Eq.~\eqref{ass-Brown},
leads to the following evolution equations~\cite{Brambilla:2017zei}:
\begin{equation}
\frac{\mathrm{d}\rho}{\mathrm{d}t}=-i[H,\rho]+\sum_{nm}h_{nm}\left(L_i^n\rho {L_i^m}^\dagger-\frac{1}{2}\{{L_i^m}^\dagger L_i^n,\rho\}\right)\,,
\end{equation}
where 
\begin{align}
\rho &=\left(\begin{array}{cc}
\rho_s & 0\\
0 & \rho_o\end{array}\right),
\label{densitymatrix}\\
H &= \left(\begin{array}{cc}
h_s & 0\\
0 & h_o 
\end{array}\right)
+ \frac{r^ir^j}{2} \tilde{\gamma}^{ij}(t)
\left(\begin{array}{cc}
1 & 0\\
0 & \tfrac{N_{c}^{2}-2}{2(N_{c}^{2}-1)}
\end{array}\right),
\\
L_i^0 &= r^i\left(\begin{array}{cc}
0 & 0 \\
0 & 1
\end{array}\right),
\\
L_i^1 &=
\frac{r^j}{2}  \left(\tilde{\kappa}^{ij}(t) + i\tilde{\gamma}^{ij}(t)\right)
\left(\begin{array}{cc}
0 & 0 \\
0 & \tfrac{N_{c}^{2}-4}{2(N_{c}^{2}-1)}
\end{array}\right),
\\
L_i^2 &= r^i \left(\begin{array}{cc}
0 & 1 \\
1 & 0
\end{array}\right),
\\
L_i^3 &= \frac{r^j}{2}  \left(\tilde{\kappa}^{ij}(t) + i\tilde{\gamma}^{ij}(t)\right)
\left(\begin{array}{cc}
0 & \frac{1}{N_c^2-1} \\
1 & 0
\end{array}\right).
\end{align}
Despite the fact that $h_{nm}$ are the elements of a matrix,
\begin{equation}
h=\left(\begin{array}{cccc}
0 & 1 & 0 & 0 \\
1 & 0 & 0 & 0 \\
0 & 0 & 0 & 1 \\
0 & 0 & 1 & 0 
\end{array}\right),
\end{equation}
which is not positive definite, it is straightforward to show that after a redefinition of $L_i^n$ 
the eigenvectors of $h$ associated to negative eigenvalues, which are proportional to $L_i^1 - L_i^0$ and $L_i^3-L_i^2$, can be set to zero. 
Hence, according to Refs.~\cite{Lindblad:1975ef,Gorini:1975nb}, we can map the above evolution equations into the Lindblad form~\eqref{eq:lindblad}.
We emphasize that the evolution equations in this form hold for a Coulombic bound state in quantum Brownian motion in a medium.
We do not require the medium to be weakly coupled.

The tensors $\tilde{\kappa}^{ij}(t)$ and $\tilde{\gamma}^{ij}(t)$ have a field theoretical definition:
\begin{alignat}{3}
\tilde{\kappa}^{ij}(t)&=&&\frac{1}{2N_{c}}\int^{t}_{t_{0}}\mathrm{d}t'\Big \langle \left\{ gE^{a,i}(t,\textbf{0}),gE^{a,j}(t',\textbf{0}) \right\} \Big \rangle,\\
\tilde{\gamma}^{ij}(t)&=-&&\frac{i}{2N_{c}}\int^{t}_{t_{0}}\mathrm{d}t'\Big \langle \left[ gE^{a,i}(t,\textbf{0}),gE^{a,j}(t',\textbf{0}) \right] \Big \rangle,
\label{gammatensor}
\end{alignat}
where $\langle \cdots \rangle$ stands for the in medium average, the curly brackets signify anticommutator, the square ones commutator, and ${\bf E}$ is the chromoelectric field.
In the above expressions, the chromoelectric field has to be understood as  $\Omega^\dagger \times (\text{usual {\bf E}}) \times \Omega$, where 
$\Omega$ is a Wilson line going from $-\infty$ to $t$: $\displaystyle \Omega = \exp\left[ -ig \int_{-\infty}^t ds \, A_0(s,{\bf 0}) \right]$.
The Wilson lines guarantee that the definitions of $\tilde{\kappa}^{ij}(t)$ and $\tilde{\gamma}^{ij}(t)$ are gauge invariant.
The tensors $\tilde{\kappa}^{ij}(t)$ and $\tilde{\gamma}^{ij}(t)$ may be related to the real and imaginary parts of the quark-antiquark color singlet 
self-energy diagram shown in Fig.~\ref{fig:sigmas}:
\begin{align}
\Sigma_{s}(t) &= r^{i}r^{j}\frac{1}{2N_{c}}\int_{t_{0}}^{t}\mathrm{d}t'\big \langle gE^{a,i}(t,\textbf{0})\,gE^{a,j}(t',\textbf{0}) \big \rangle 
\nonumber\\
&=\frac{r^ir^j}{2}\left[ \tilde{\kappa}^{ij}(t)+i\tilde{\gamma}^{ij}(t) \right].
\label{eq:exact_selfenergy}
\end{align}

If the medium is {\em isotropic}, then $\tilde{\kappa}^{ij}(t) = \delta^{ij}\tilde{\kappa}(t)$ and $\tilde{\gamma}^{ij}(t) = \delta^{ij}\tilde{\gamma}(t)$, where 
\begin{alignat}{3}
\tilde{\kappa}(t)&=&&\frac{1}{6N_{c}}\int^{t}_{t_{0}}\mathrm{d}t'\Big \langle \left\{ gE^{a,i}(t,\textbf{0}),gE^{a,i}(t',\textbf{0}) \right\} \Big \rangle,
\label{kappat}\\
\tilde{\gamma}(t)&=-&&\frac{i}{6N_{c}}\int^{t}_{t_{0}}\mathrm{d}t'\Big \langle \left[ gE^{a,i}(t,\textbf{0}),gE^{a,i}(t',\textbf{0}) \right] \Big \rangle.
\label{gammat}
\end{alignat}
The Lindblad equation describing the time evolution of the density matrix 
has then a particularly simple form~\cite{Brambilla:2016wgg,Brambilla:2017zei}, as the Hermitian operator $H$ is given by 
\begin{align}
H&=\begin{pmatrix} h_{s} & 0 \\ 0 & h_{o} \end{pmatrix}
+\frac{r^{2}}{2}\tilde{\gamma}(t)\begin{pmatrix} 1 & 0 \\ 0 & \tfrac{N_{c}^{2}-2}{2(N_{c}^{2}-1)} \end{pmatrix}, \label{eq:hamiltonian} 
\end{align}
and the collapse operators $C_{i}$ by:
\begin{align}
C_{i}^{0}&=\sqrt{\frac{\tilde{\kappa}(t)}{N_{c}^{2}-1}}r^{i}\begin{pmatrix} 0 & 1 \\ \sqrt{N_{c}^{2}-1} & 0 \end{pmatrix}, \label{eq:collapse0} \\
C_{i}^{1}&=\sqrt{\frac{(N_{c}^{2}-4)\tilde{\kappa}(t)}{2(N_{c}^{2}-1)}}r^{i}\begin{pmatrix} 0 & 0 \\ 0 & 1 \end{pmatrix}. \label{eq:collapse1}
\end{align}

At short times after the formation of the medium,  $t \gtrsim t_{0}$, $\tilde{\kappa}(t)$ scales like $T^{4}(t-t_{0})$ and the thermal part of $\tilde{\gamma}(t)$ like $T^{5}(t-t_{0})^2$.
In the opposite, large time limit, $t-t_{0}$ is the largest time scale in the problem.
At this point, it is convenient to assume that the chromoelectric correlators appearing in \eqref{kappat} and \eqref{gammat} 
are, at least approximately, {\em time translation invariant}:
$ \langle  E^{a,i}(t,\textbf{0}) E^{a,i}(t',\textbf{0})  \rangle = \langle  E^{a,i}(t-t',\textbf{0}) E^{a,i}(0,\textbf{0})  \rangle$.
This is the case, for instance, at thermal equilibrium or close to it, if the variation in time of the temperature is slow.
In the large time limit and under the assumption of (approximate) time translation invariance $\tilde{\kappa}(t)$ and  $\tilde{\gamma}(t)$ approach the 
asymptotic values $\kappa=\tilde{\kappa}(\infty)$ and $\gamma=\tilde{\gamma}(\infty)$ respectively:
\begin{alignat}{3}
\kappa&=&&\frac{1}{6N_{c}}\int^{\infty}_{0}\mathrm{d}t~\Big \langle \left\{ gE^{a,i}(t,{\bf 0}),gE^{a,i}(0,{\bf 0}) \right\} \Big \rangle, \label{eq:kappa_def} \\
\gamma&=-&&\frac{i}{6N_{c}}\int^{\infty}_{0}\mathrm{d}t~\Big \langle \left[ gE^{a,i}(t,{\bf 0}),gE^{a,i}(0,{\bf 0}) \right] \Big \rangle. \label{eq:gamma_def}
\end{alignat}
The quantity $\kappa$ is the heavy quark momentum diffusion coefficient.
The above form of the heavy quark momentum diffusion coefficient was first derived in~\cite{CasalderreySolana:2006rq} (see also~\cite{CaronHuot:2009uh})
in the context of the diffusion of a heavy quark in a thermal medium according to the Langevin equations~\eqref{eq:langevin}.
The coefficient $\kappa$ can also be written as the real part of the time ordered correlator 
${1}/{(6\,N_c)} \int_{-\infty}^{+\infty}\mathrm{d}t \, \langle T\,gE^{a,i}(t,{\bf 0}) \,gE^{a,i}(0,{\bf 0})\rangle$; $\gamma$ is then its imaginary part.

In the large time limit, for an isotropic medium and under approximate time translation invariance, the color singlet self energy \eqref{eq:exact_selfenergy} becomes
\begin{equation}
\Sigma_{s}=\frac{r^{2}}{2}(\kappa+i\gamma). 
\label{eq:sigma}
\end{equation}
This allows to write 
\begin{alignat}{3}
r^2\kappa &= \Sigma_{s}+\Sigma^{\dagger}_{s} && = -2\,\mathrm{Im}(-i\Sigma_{s}), \label{eq:Gamma}\\
r^2\gamma &= -i\Sigma_{s}+i\Sigma_{s}^{\dagger} && = ~~\, 2\,\mathrm{Re}(-i\Sigma_{s}), \label{eq:deltam}
\end{alignat}
and eventually relate $\kappa$ and $\gamma$ to the quarkonium in medium width, $\Gamma$, and the in medium mass shift, $\delta M$.
For $1S$ Coulombic quarkonium states, these relations read~\cite{Brambilla:2016wgg,Brambilla:2017zei}
\begin{align}
\Gamma(1S) &= 3a_{0}^{2}\kappa, \label{eq:kappa} \\
\delta M(1S) &= \frac{3}{2}a_{0}^{2}\gamma \label{eq:gamma},
\end{align}
where 3$a_{0}^{2}$ is the expectation value of $r^2$ on a $1S$ Coulombic bound state.
The Bohr radius is $a_{0}=2/(MC_{F}\alpha_{s})$.

$\Gamma$ and $\delta M$ do not contain, by definition, in vacuum contributions.
Also $\kappa$, as defined in~\eqref{eq:kappa_def}, does not contain in vacuum contributions, reflecting the fact that  
energy conservation prohibits the decay of a heavy quark-antiquark color singlet into a heavy quark-antiquark color octet in vacuum.
In contrast, $\gamma$, as defined in~\eqref{eq:gamma_def}, does contain in vacuum contributions.
Hence, Eq.~\eqref{eq:gamma} relates $\delta M(1S)$ to $\gamma$ subtracted of its vacuum ($T=0$) part.
The coefficient $\gamma$ should be understood in this subtraction scheme in Eq.~\eqref{eq:gamma} and in the next section.
Now that we have explicit relations for $\kappa$ and $\gamma$ in terms of $a_{0}$, $\Gamma(1S)$, and $\delta M(1S)$, 
we can proceed to extract $\kappa$ and $\gamma$ from available lattice estimates of $\Gamma(1S)$ and $\delta M(1S)$.

\section{Results and comparisons}
\label{sec:numerical_calculations}
Equations~\eqref{eq:kappa} and \eqref{eq:gamma} fix the ratio $\gamma/\kappa$ to be 
\begin{equation}
\frac{\gamma}{\kappa} = 2 \,\frac{\delta M(1S)}{\Gamma(1S)},
\end{equation}
which may turn out to be useful once both $\delta M(1S)$ and $\Gamma(1S)$ are reliably determined, for the ratio does not depend on the Bohr radius. 
The quantities $\delta M(1S)$ and $\Gamma(1S)$ cannot be accessed by experiments, as in heavy-ion collisions the quarkonium decays, at a late time, in the vacuum.
These quantities can instead by computed by lattice QCD, with the thermal mass shift, $\delta M(1S)$, clearly in a more reliable fashion than the thermal width, $\Gamma(1S)$. 

In order to determine $\gamma$ and $\kappa$ from  $\delta M(1S)$ and $\Gamma(1S)$ using Eqs.~\eqref{eq:gamma} and \eqref{eq:kappa}, we need to calculate $a_{0}$. 
Since the system is Coulombic, we can do it by solving the self-consistency equation
\begin{equation}
a_{0}=\frac{2}{MC_{F}\alpha_{s}(1/a_{0})},
\label{self-cons-a0}
\end{equation}
where $\alpha_{s}(1/a_{0})$ is the strong coupling evaluated at the scale $1/a_0$. 
For the bottom and charm masses, we take $M=M_{b}=4.78$~GeV and $M=M_{c}=1.67$~GeV, respectively. 
These are the central values for the pole masses quoted by the Particle Data Group~\cite{Tanabashi:2018oca}.\footnote{\label{foot2}
Since in Ref.~\cite{Kim:2018yhk} the masses $M_{b}=4.65$~GeV and $M_{c}=1.275$~GeV were used, 
we have rescaled their values for $\delta M(1S)$ and $\Gamma(1S)$ by $(4.65/4.78)^2$ in the bottomonium case and by $(1.275/1.67)^2$ in the charmonium case. 
Similarly, since in Ref.~\cite{Aarts:2011sm} the mass $M_{b}=5$~GeV was used, we have rescaled their value for $\Gamma(1S)$ by $(5/4.78)^2$.
The coefficients $\gamma$ and $\kappa$ are mass independent, so the choice of the mass should not affect them.
There is, however, a residual dependence due to having truncated the expressions in the right-hand sides of Eqs.~\eqref{eq:kappa} and~\eqref{eq:gamma} at 
leading order in the various expansions underlying the effective field theory. 
We have checked that this residual dependence is, indeed, well accounted for by the quoted errors.
}
We solve for $a_{0}$ using the one-loop, 3-flavor running of $\alpha_{s}$ with $\Lambda_{\overline{\text{MS}}} = 332$~MeV, also from Ref.~\cite{Tanabashi:2018oca}.  
We take the running of $\alpha_s$ at one loop for consistency with the fact that the radius of a $1S$ Coulombic bound state is given by Eq.~\eqref{self-cons-a0} only at leading order.
For the bottomonium ground state we obtain $a_{0} = 0.67\text{ GeV}^{-1} = 0.13\text{ fm}$, 
while for the charmonium ground state we obtain $a_{0} = 1.19\text{ GeV}^{-1} = 0.23\text{ fm}$.
For the above choice of heavy quark masses, the $\Upsilon(1S)$ binding energy is $E = M(\Upsilon(1S)) - 2M_b = - 0.1\text{ GeV}$, 
and the $J/\psi$ binding energy is $E = M(J/\psi) - 2M_c = - 0.24\text{ GeV}$.\footnote{Defining the binding energy as the Coulombic Bohr level, $E =-1/(M a_0^2)$, changes the numerical value (to $E = -0.46$~GeV for the $\Upsilon(1S)$ and $E = -0.42$~GeV for the $J/\psi$), but not the hierarchy of energy scales. Hence, the following arguments and extractions of $\gamma$ and $\kappa$, which only depend on that hierarchy, would remain unchanged.}

\begin{figure}[ht]
\includegraphics[width=\linewidth]{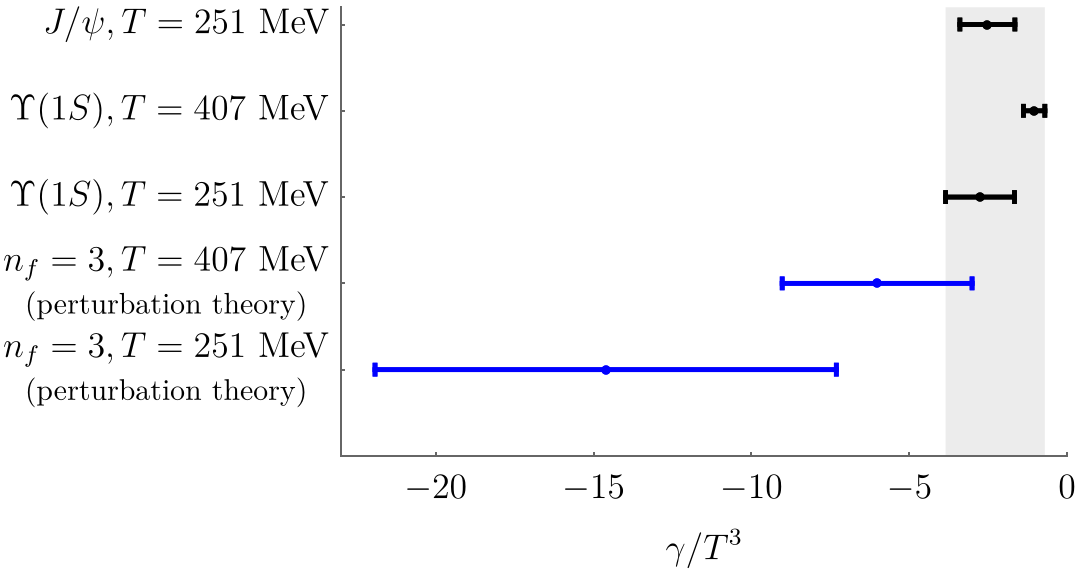}
\caption{The first three entries (black bars) show $\gamma/T^3$ as obtained from Eq.~\eqref{eq:gamma} using lattice data of Ref.~\cite{Kim:2018yhk}
for the thermal mass shift of the $J/\psi$ and of the $\Upsilon(1S)$ at two different temperatures. The error bars account for the lattice uncertainties only.
The last two entries (blue bars) provide  $\gamma/T^3$ from the perturbative, leading order, expression of the thermal mass shift
with the strong coupling computed at $\pi$ times the two different temperatures $251\text{ MeV}$ and $407\text{ MeV}$.
We assign a 50\% uncertainty to these results. The gray band gives our final range for $\gamma/T^3$, see text.}
\label{fig:gamma_estimates}
\end{figure}

We evaluate $\gamma/T^3$ using the thermal mass shifts computed for the $J/\psi$ at $T=251\text{ MeV}$ and for the $\Upsilon(1S)$ at $T=407\text{ MeV}$ and $T=251\text{ MeV}$ 
in the 2+1 flavor lattice simulation of Ref.~\cite{Kim:2018yhk}. 
After rescaling for the mass, see Footnote~\ref{foot2}, the mass shifts are: 
$(-85\pm 29)\text{ MeV}$ for the $J/\psi$ at $T=251\text{ MeV}$, and  $(-48\pm 16)\text{ MeV}$ and  $(-30\pm 12)\text{ MeV}$ for the $\Upsilon(1S)$ at $T=407\text{ MeV}$ and $T=251\text{ MeV}$, respectively.\footnote{
Consistently with the Coulombic assumption, the $J/\psi$ and $\Upsilon(1S)$ in vacuum binding energies are negative 
and at least a factor 2 larger than the corresponding thermal mass shifts.
}
The results for $\gamma$ are shown by the three first entries (black bars) of Fig.~\ref{fig:gamma_estimates}.
For the $J/\psi$, we have $1/a_{0}=0.84\text{ GeV}$, which fulfills (somewhat marginally)
the hierarchy of Eq.~\eqref{ass-coul} at $T=251\text{ MeV}$, since $\pi \times (251\text{ MeV}) = 0.79\text{ GeV}$. 
It also fulfills the condition~\eqref{ass-Brown}.
In the case of the $\Upsilon(1S)$, $1/a_{0}=1.5\text{ GeV}$ and both conditions~\eqref{ass-coul} and~\eqref{ass-Brown}
are fulfilled at both temperatures $T=251\text{ MeV}$ and $T=407\text{ MeV}$ although at the lower temperature more clearly than at the higher one. 
The two conditions~\eqref{ass-coul} and~\eqref{ass-Brown} guarantee that the $J/\psi$ and $\Upsilon(1S)$ remain Coulombic also in the medium 
and that their motion through the medium is a quantum Brownian one.
Because of this, we consider all three extractions of $\gamma$, from the $J/\psi$ (one temperature) and the $\Upsilon(1S)$ (two temperatures), reliable, 
as they are consistent with our assumptions.
We take their range as an estimate of $\gamma/T^3$ for $251\text{ MeV}  \lesssim T \lesssim 407\text{ MeV}$:
\begin{equation}
-3.8 \lesssim \frac{\gamma}{T^{3}} \lesssim -0.7\,.
\label{finalgamma}
\end{equation}
It is particularly significant to see that the extraction of $\gamma/T^3$ from the $J/\psi$ at $T=251\text{ MeV}$ overlaps perfectly with the 
extraction of $\gamma/T^3$ from the $\Upsilon(1S)$ at the same temperature. This shows that, as expected, $\gamma/T^3$ depends only on the temperature, 
while it does not depend on the quarkonium state. 
Concerning the temperature dependence, the extraction of $\gamma/T^3$ from the $\Upsilon(1S)$ at $T=407\text{ MeV}$ could suggest that 
$-\gamma/T^3$ tends towards smaller values at higher temperatures.
The last two entries (blue bars) of Fig.~\ref{fig:gamma_estimates} refer to $\gamma$ determined from the leading order expression of the thermal mass shift computed in Ref.~\cite{Brambilla:2008cx} 
and reported in Refs.~\cite{Brambilla:2016wgg,Brambilla:2017zei}. 
We see that perturbation theory also gives a negative value for $\gamma$, as our non-perturbative estimate above. 
Moreover, there is a partial overlap between the perturbative result at the highest temperature, 
where a weak-coupling treatment is expected to work better, and the range given in Eq.~\eqref{finalgamma}.
It is possible, however, that higher-order corrections will spoil the leading order result at the temperatures considered here, as it is the case for the 
weak-coupling expression of~$\kappa$.
Finally, we remark that a small and negative value of $\gamma/T^3$ is also phenomenologically favored by the comparison of the $\Upsilon(1S)$ nuclear modification factor, 
as computed from the Lindblad equation \eqref{eq:lindblad}, \eqref{eq:hamiltonian}, \eqref{eq:collapse0}, \eqref{eq:collapse1}, 
with the most recent CMS data~\cite{Brambilla:2016wgg,Brambilla:2017zei,Espinosa:2018dfq}. 

\begin{figure}[ht]
\includegraphics[width=\linewidth]{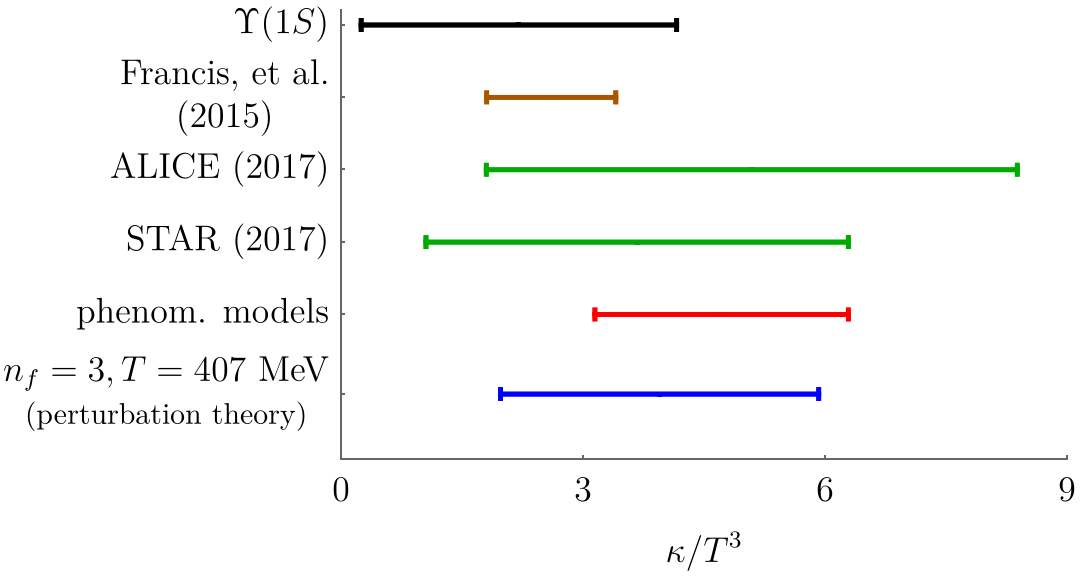}
\caption{The first entry (black bar) shows $\kappa/T^3$ as obtained from Eq.~\eqref{eq:kappa} using lattice data of Refs.~\cite{Aarts:2011sm,Kim:2018yhk} 
for the upper and lower bounds of the thermal decay width of the $\Upsilon(1S)$.
The second entry (brown bar) reports the (quenched) lattice estimate of Ref.~\cite{Francis:2015daa}.
The third and fourth entries (green bars) are the determinations based on the ALICE~\cite{Acharya:2017qps} and STAR~\cite{Adamczyk:2017xur}
measurements of the $D$-meson azimuthal anisotropy coefficient $v_{2}$, respectively.
The fifth entry (red bar) is the determination based on a range of values of the heavy quark spatial diffusion coefficient obtained in \cite{Dong:2019byy} from an analysis of phenomenological models.
The sixth entry (blue bar) is a perturbative result (see text) 
with the strong coupling computed at the scale  $\pi \times (407\text{ MeV}) = 1.28\text{ GeV}$. We assign a 50\% uncertainty to it.}
\label{fig:kappa_estimates}
\end{figure}

Concerning $\kappa$, the available lattice data for the thermal width are less precise than those available for the mass shift.  
The width given in the 2+1 flavor lattice simulation of Ref.~\cite{Kim:2018yhk} is preliminary and should be understood 
as a lower bound rather than the width itself~\cite{privatecommunication}.
We combine this lower bound with the somewhat older 2 flavor lattice results from~\cite{Aarts:2011sm} that supply an upper bound to the thermal width.  
More specifically, we take the $\Upsilon(1S)$ width at T=407 MeV from~\cite{Kim:2018yhk} ($\Gamma(1S) \approx 22.3\text{ MeV}$, after rescaling for the mass) as a lower bound 
and the highest temperature estimate of $\Gamma(1S)$ from Fig.~5 of Ref.~\cite{Aarts:2011sm} 
($\Gamma(1S)/T \approx 1.1$, after rescaling for the mass, for $T/T_{c} \approx 2$ with $T_{c} \approx 220\text{ MeV}$) as an upper bound.  
We obtain
\begin{equation}
0.24 \lesssim \frac{\kappa}{T^{3}} \lesssim 4.2\,.
\label{finalkappa}
\end{equation}
The above range is the first entry (black bar) in Fig.~\ref{fig:kappa_estimates}.
We note that we do not plot an estimate of $\kappa$ obtained from the thermal width of the $J/\psi$ as Ref.~\cite{Aarts:2011sm} performed measurements only on bottomonium states; this leaves us with only a lower bound from the measurement of the width of the $J/\psi$ at $T=251$ MeV performed in Ref.~\cite{Kim:2018yhk}. 
We find this gives a lower limit of $\kappa/T^{3}\gtrsim 0.235\pm 0.208$ (with purely statistical uncertainties); we note the agreement with the lower bound of $\kappa/T^{3}\gtrsim0.24$ obtained from the $\Upsilon(1S)$ at T=407 MeV.
The second entry (brown bar) in Fig.~\ref{fig:kappa_estimates} reports the result, $1.8 \lesssim \kappa/T^3\lesssim 3.4$,  
of the lattice study done in a pure SU(3) plasma at $T\approx 1.5\,T_{c}$ with $T_c \approx 313\text{ MeV}$ in Ref.~\cite{Francis:2015daa}.  
Indirect bounds can also be placed on $\kappa$ from the experimental measure of the $D$-meson azimuthal anisotropy coefficient $v_{2}$.  
In particular, measurements from the ALICE~\cite{Acharya:2017qps} and STAR~\cite{Adamczyk:2017xur} collaborations can be compared 
with theoretical models to place bounds on the heavy quark spatial diffusion coefficient.  
Relating the heavy quark spatial diffusion coefficient $D$ to the heavy quark momentum diffusion coefficient ($\kappa/T^{3}=2/DT$), 
we find that ALICE data give $1.8 \lesssim \kappa/T^{3} \lesssim 8.4$ at $T=T_c$ and STAR data give $1.0 \lesssim \kappa/T^{3} \lesssim 6.3$ for $T_c \lesssim T \lesssim 2\,T_c$. 
The third and fourth entries (green bars) in Fig.~\ref{fig:kappa_estimates} represent these two bounds, respectively.
Calculations of the nuclear modification factor $R_{AA}$ and the elliptic flow $v_{2}$ from phenomenological models also place bounds on the momentum diffusion coefficient; for a collection of recent results, see \cite{Dong:2019byy}. 
The results of this reference give $3.1\lesssim \kappa/T^{3}\lesssim 6.3 $ at $T\approx 155-160$ MeV and are the fifth entry (red bar) in Fig.~\ref{fig:kappa_estimates}.\footnote{We note that the phenomenological models used in the last three extractions may be more complicated than the Langevin equation given in Eq.~\ref{eq:langevin}. Furthermore, these extractions are not based on a non-relativistic expansion and may include into $\kappa$ dynamics that in our effective field theory framework occur as relativistic corrections of higher order in $1/M$; these corrections scale as the square of the relative heavy-quark velocity in the quarkonium possibly contributing to a systematic effect by up to $10\%$ for bottomonium and $30\%$ for charmonium when comparing with the mass-independent extractions from the chromoelectric correlator.}
The sixth entry (blue bar) in Fig.~\ref{fig:kappa_estimates} follows from computing $\kappa$ in weak-coupling perturbation theory at $T = 407\text{ MeV}$.
In the perturbative expression we include the complete order $g^4$ contribution and the order $g^5$ term $7/(48\pi^2)\,C_F N_c g^4 (m_D/T)$
(this is, truncated at order $g^5$, what is called leading order expression in~\cite{CaronHuot:2008uh}); we note that the order $g^4$ contribution alone would give unphysical negative values of $\kappa$ for realistic couplings. 
The perturbative expression of $\kappa/T^3$, in the above sense, gives $2 \lesssim \kappa/T^3 \lesssim 6$ by assigning a 50\% uncertainty. 
This uncertainty may, however, be underestimated since the complete $g^5$ correction is known and very large~\cite{CaronHuot:2008uh}. 
In fact it may increase the leading order result even by an order of magnitude under some circumstances~\cite{CaronHuot:2008uh,Banerjee:2011ra}, 
which obviously questions the reliability of a naive weak-coupling expansion for $\kappa$ at the considered temperatures.

All determinations of $\kappa/T^3$ are consistent with each other, and, in particular, with the range presented in Eq.~\eqref{finalkappa}.
This is noteworthy as these determinations are very different, some of them do not even rely on full QCD, and have been obtained at different temperatures.
Indeed, $\kappa/T^3$, as well as $\gamma/T^3$, does depend on the temperature.
The lattice study of~\cite{Banerjee:2011ra} could suggest $\kappa/T^3$ assumes lower values at higher temperatures, which seems consistent with studies from phenomenological models \cite{Das:2015ana}.
Finally, we mention that the same relationship between the width of the state and $\kappa$ given by Eq.~\eqref{eq:kappa} 
has also been exploited in~\cite{Datta:2018ltb}, but in the opposite direction, to supply an estimate of the quarkonium thermal width from~$\kappa$.

\section{Conclusion}
\label{sec:conclusion}
The Lindblad equation describing the heavy quarkonium evolution in the hot medium created at the early stages 
of high-energy heavy-ion collisions requires, under some conditions, only two parameters 
to describe the interaction of the heavy quark-antiquark pair with the medium~\cite{Brambilla:2016wgg,Brambilla:2017zei,us:2018}.
The conditions are that the quarkonium is a Coulombic bound state, which holds when Eq.~\eqref{ass-coul} or Eq.~\eqref{RggE} are satisfied, 
and that its motion in the medium is a quantum Brownian motion, which holds when Eq.~\eqref{ass-Brown} or Eq.~\eqref{RggS} are satisfied.
The two parameters are the heavy quark momentum diffusion coefficient, $\kappa$, which crucially enters also the Langevin equation describing
the heavy quark diffusion in the medium, whose field theoretical definition is in Eq.~\eqref{eq:kappa_def}, and its dissipative counterpart, 
the coefficient $\gamma$, whose field theoretical definition is in Eq.~\eqref{eq:gamma_def}.
No assumption is required on the nature of the medium.

In this paper, we have estimated $\gamma$, using its relation to the in medium quarkonium mass shift and the 2+1 flavor lattice data of Ref.~\cite{Kim:2018yhk}.
The result is given in Eq.~\eqref{finalgamma}. This is the first non-perturbative determination of $\gamma$. 
Its sign is consistent with the weak-coupling, leading order thermal mass shift, which, however, is affected by large uncertainties.

We have also computed $\kappa$, using its relation to the in medium quarkonium decay width and the two different sets of lattice data~\cite{Aarts:2011sm,Kim:2018yhk}.
The data are not precise enough to pin down a narrow range of $\kappa$, nevertheless they allow us to establish an upper and a lower limit for this transport coefficient. 
They are given in Eq.~\eqref{finalkappa}. The range of $\kappa$ is consistent with other determinations, see Fig.~\ref{fig:kappa_estimates}.
Once more precise lattice data will become available, this method has the potential to provide a competitive determination of the heavy quark momentum diffusion coefficient. 
Already now, it gives a range for $\kappa$ that is based on 2+1 and 2 flavor lattice data, while current lattice determinations rely on pure SU(3) gauge theory simulations~\cite{Banerjee:2011ra,Francis:2015daa}.

From the above considerations, it is clear that this work calls for several further lattice analyses.
Concerning $\gamma$, it would be important to have a direct evaluation based on the spectral function of the chromoelectric field correlator and Eq.~\eqref{gamma-spectral} in a given renormalization scheme.
Concerning $\kappa$, besides direct full QCD determinations based on Eq.~\eqref{kappa-spectral}, 
this work aims also at motivating further lattice computations of the in medium quarkonium decay width.

Finally, we remark that the logic of the present work may be reversed and the agreement noticed between the present determination of $\kappa$ and 
other independent ones used to support our starting assumptions on the nature of the studied heavy quarkonia and of their diffusion in the medium.
Indeed, our determination of $\kappa$ from the in medium decay width and the explicit computation of the relevant energy scales 
provide evidence that at least the $\Upsilon(1S)$ is a Coulombic bound state that propagates with a quantum Brownian motion 
in the medium formed, and at the temperature attained, by present day heavy-ion colliders.

\section*{Acknowledgements}
The authors thank Jacopo Ghiglieri, Viljami Leino, Peter Petreczky and Alexander Rothkopf for discussions and a reading of the manuscript. 
N.B. and A.V. thank Yukinao Akamatsu and Joan Soto for an inspiring discussion at the Institute for Nuclear Theory while attending the program 
``Multi-Scale Problems Using Effective Field Theories''.
This work was funded by the \textit{Bundesministerium f\"{u}r Bildung und Forschung} project no. 05P2018 
and by the DFG cluster of excellence ``Universe'' (www.universe-cluster.de).  
The work of M.A.E. was supported by the Academy of Finland project 297058, by \textit{Ministerio de Ciencia e Innovacion} of Spain 
under project FPA2017-83814-P and Maria de Maetzu Unit of Excellence MDM-2016- 0692, by Xunta de Galicia and FEDER.  
We thank the Munich Institute for Astro- and Particle Physics (MIAPP) of the DFG cluster of excellence ``Universe'' 
during which activity ``Probing the Quark-Gluon Plasma with Collective Phenomena and Heavy Quarks'' the idea of this paper was first discussed.

\appendix
\section{Spectral function and $\gamma$}
\label{appendix}
In this appendix, we define the spectral function of the chromoelectric correlator, $\rho_{\text{el}}$, 
and derive expression \eqref{gamma-spectral} for $\gamma$.

We start considering the real time chromoelectric correlator
\begin{equation}
\Big\langle gE^{a,i}(t,{\bf 0}) \, gE^{a,i}(0,{\bf 0}) \Big\rangle,
\label{Appcorr}
\end{equation}
where the chromoelectric fields are understood in the convention specified after Eq.~\eqref{gammatensor}, i.e., with Wilson lines attached.
The Wilson lines make the correlator gauge invariant.
The medium average $\langle \cdots \rangle$ is normalized by the partition function.
In turn, we normalize the partition function in such a way that, for vanishing coupling, it agrees with the partition function of the Abelian theory.
This normalization differs by a factor $N_c$ from a common choice in the literature.

The {\em retarded} chromoelectric correlator is defined as
\begin{equation}
G_{\text{R}}(t) = \theta(t) \Big\langle \left[ gE^{a,i}(t,{\bf 0}), gE^{a,i}(0,{\bf 0}) \right]\Big\rangle,
\label{Appret}
\end{equation}
and its Fourier transform as
\begin{equation}
G_{\text{R}}(\omega) = \int \mathrm{d}t\,e^{i\omega t}\,G_{\text{R}}(t).
\label{AppretF}
\end{equation}
Following Ref.~\cite{CasalderreySolana:2006rq}, the spectral function of the chromoelectric correlator is defined as
\begin{equation}
\rho_{\text{el}}(\omega) = 2 \,\text{Im}\left[ i G_{\text{R}}(\omega) \right].
\label{Apprho}
\end{equation}
From this definition and the definition of $\kappa$, we get Eq.~\eqref{kappa-spectral}.
Equation~\eqref{kappa-spectral} agrees with expressions found in the literature.
In particular, it agrees with the expression of Ref.~\cite{CasalderreySolana:2006rq} by taking into account that
the spectral function of~\cite{CasalderreySolana:2006rq} differs by $1/N_c \times 1/2 \times 1/3$ from ours,
where the first factor comes from the different normalization in color of the partition function,
the second one from the trace over the color matrices of the chromoelectric fields, $\text{Tr}\{T^a T^b\} = \delta^{ab}/2$,
and the third one from the average over the spatial directions.
It also agrees with the expression found in~\cite{Burnier:2010rp,Banerjee:2011ra,Francis:2015daa},
if one takes into account a further factor $1/2$ difference in the definition of the spectral function.

The leading order weak-coupling expression of the in vacuum spectral function is
\begin{equation}
\rho_{\text{el}}(\omega) = g^2\,(N_c^2-1) \, \frac{\omega^3}{\pi}\,,
\label{ApprhoLO}
\end{equation}
which, once converted, agrees with~\cite{Burnier:2010rp}. Higher-order corrections add powers of $g^2$, but do not modify the functional behaviour in $\omega$, 
since $\omega$ is the only scale in the vacuum.
Equation \eqref{ApprhoLO} describes the large frequencies behaviour of the spectral function, as at large frequencies all energy scales other than $\omega$ can be ignored.

From the definition of $\gamma$, given in Eq.~\eqref{eq:gamma_def}, it results that
\begin{equation}
\gamma = -\frac{1}{6N_c}  iG_{\text{R}}(\omega=0).
\label{gammaretarded}
\end{equation}
We can relate $iG_{\text{R}}(\omega)$ to its imaginary part using a dispersion relation.
If no subtraction is needed, it reads
\begin{equation}
iG_{\text{R}}(\omega) = \int  \frac{\mathrm{d}\omega'}{2\pi} \frac{\rho_{\text{el}}(\omega')}{\omega' - \omega -i\eta}. 
\end{equation}
Substituting this into Eq.~\eqref{gammaretarded}, we obtain Eq.~\eqref{gamma-spectral}. 
Note that, since $\rho_{\text{el}}(\omega)$ vanishes for $\omega=0$, there is no imaginary contribution to $\gamma$ from the dispersion relation.
Moreover, since $\rho_{\text{el}}(\omega)/\omega$ is an even function in $\omega$, we could write the integral over the real axis as twice the integral over the positive real axis.

\medskip

{\em Note added:} While finishing this paper, we became aware of the preprint of A.~M.~Eller, J.~Ghiglieri and G.~D.~Moore 
``Thermal quarkonium mass shift from Euclidean correlators''~\cite{Eller:2019spw} that proposes a way to determine $\gamma$ directly from a proper Euclidean correlator.
We thank the authors for sharing with us their results prior of publication.

\pagebreak

\bibliographystyle{apsrev4-1} 
\bibliography{kappa_gamma_bibliography} 

\end{document}